\documentstyle[twocolumn,aps,prd,epsf]{revtex}
\begin{document}
\draft
\title{Analytic proof of the Weinberg theorem 
for off mass shell and composite pions}
\author{
Poster for the 5th ''Quark confinement and the hadron spectrum'' Conference, 
Gargnano, September 2002
\\
Pedro Bicudo
}
\address{
Departamento de F\'\i sica, and CFIF, Instituto Superior T\'ecnico,
Av. Rovisco Pais, 1049-001 Lisboa, Portugal
}
\maketitle
\begin{abstract} 
The Weinberg theorem for the pion-pion 
scattering is proved analytically in full detail 
for quark models in the ladder truncation.
The proof is displayed with Feynman diagrams.
The axial and vector Ward identities, 
for the quark propagator and for the ladder, 
exactly cancel any model dependence.
Off mass shell and finite size effects are
included in the quark-antiquark pion 
Bethe Salpeter vertices. 
This is applied to compute the isospin
0, 1 and 2 scattering matrices 
off the mass shell .
\end{abstract}
\pacs{ }
\par
%sssssssssssssssssssssssssssssssssssssssssssssssssssssssssssssssssssssssss
%sssssssssssssssssssssssssssssssssssssssssssssssssssssssssssssssssssssssss
%sssssssssssssssssssssssssssssssssssssssssssssssssssssssssssssssssssssssss
%sssssssssssssssssssssssssssssssssssssssssssssssssssssssssssssssssssssssss
%
\vspace{-.5cm}
\par
The success of the Quark Model 
\cite{Rujula}
relies on its ability to reproduce
the Hadron spectrum with microscopical interacting quarks. This
includes for instance the linear Regge trajectories, 
and the correct spin-spin and spin-orbit interactions. Moreover
the quark model is competent to predict microscopically the
hadron-hadron interactions
\cite{Ribeiro}.
For recent coupled channel studies see
\cite{Gastao,Goncalo,pi-pi}. 
However the quark model suffered
from initial difficulties 
of fully understanding the low pion mass
and the small pi-pi scattering length.
With the aim to cure this important problem, chiral symmetry breaking 
was introduced in the quark model.
\cite{Pene,Adler,Bicudo}. 
This opened the possibility to extend the 
low energy pion theorems of current algebra
to the quark model framework. 

\par
In this paper we prove that the $\pi-\pi$ scattering theorem 
of Weinberg 
\cite{Weinberg} 
applies to quark models with chiral 
invariant quark-quark interactions,
including the Nambu and Jona-Lasinio model 
\cite{Nambu,Veronique},
conventional quark models with confining 
instantaneous interactions 
\cite{Goncalo,pi-pi}, 
models with Euclidean space integrations
\cite{Liu,pi-pi,Steve},
and also covariant models in Minkowsky space
\cite{Sauli}.
After the original work of Weinberg
\cite{Weinberg}, 
the theorem was also derived 
with Ward identities for the pion fields 
\cite{Leutwyler} 
and  with a functional integration of quarks
\cite{Roberts}.
However these methods use effective pions, 
whereas pions in quark models
have naturally a finite size 
\cite{pion size}
which might affect $\pi-\pi$ scattering. 
\par
Here we complete in full detail an 
analytical proof which was recently outlined 
in \cite{pi-pi}, 
using Ward identities and the diagrammatic method  
\cite{scalar}. 
Moreover we extend the proof to 
pions with off mass shell momenta.
Amplitudes for off mass shell pions are crucial 
because $\pi-\pi$ scattering is experimentally 
estimated from $\pi-N$ scattering. The 
nucleon $N$ provides a virtual off mass shell pion $\pi^*$
that scatters with the incident $\pi$. 
For instance a virtual pion $\pi^*$ with offshellness 
$P^2-M_\pi^2=-3.32 M_\pi^2$ is present in the 
$\pi$ production at threshold.

\par
To prove the Weinberg theorem in the Quark Model 
formalism we have to truncate the series of Feynman 
diagrams consistently.
The ladder truncation is commonly assumed
for the boundstate equations,
%
%LADDER
%
\begin{equation}
%
%>>>>>>>>>>>>>>>>>>>>>>>>>>>>>>>>>>>>>>>>>>>>>>>>>>>>>>>>>>>>>>
\begin{picture}(50,20)(0,0) 
\put(0,-5){
%\put(0,0){\framebox(50,20){}}
%
\put(0,5){                         
\begin{picture}(25,20)(0,0)
%%%%%%%%%%%%%%%%%%%%%%%%%%%%% left propagators + box
\put(0,15){\line(1,0){5}}
\put(0,5){\vector(1,0){10}}
\put(15,15){\vector(-1,0){10}}
\put(15,5){\line(-1,0){5}}
\put(15,0){\framebox(10,20){}}
\end{picture}}
\put(25,5){                         
\begin{picture}(15,20)(0,0)
%%%%%%%%%%%%%%%%%%%%%%%%%%%%% right propagators
\put(0,15){\line(1,0){5}}
\put(0,5){\vector(1,0){10}}
\put(15,15){\vector(-1,0){10}}
\put(15,5){\line(-1,0){5}}
\end{picture}}
}
\end{picture}
%<<<<<<<<<<<<<<<<<<<<<<<<<<<<<<<<<<<<<<<<<<<<<<<<<<<<<<<<<<<<<
=
%>>>>>>>>>>>>>>>>>>>>>>>>>>>>>>>>>>>>>>>>>>>>>>>>>>>>>>>>>>>>>>
\begin{picture}(25,15)(0,0) 
\put(0,-5){
%\put(0,0){\framebox(40,15){}}
%
\put(0,5){                         
\begin{picture}(15,20)(0,0)
%%%%%%%%%%%%%%%%%%%%%%%%%%%%% right propagators
\put(0,15){\line(1,0){5}}
\put(0,5){\vector(1,0){10}}
\put(15,15){\vector(-1,0){10}}
\put(15,5){\line(-1,0){5}}
\end{picture}}
}
\end{picture}
%<<<<<<<<<<<<<<<<<<<<<<<<<<<<<<<<<<<<<<<<<<<<<<<<<<<<<<<<<<<<<<
+
%>>>>>>>>>>>>>>>>>>>>>>>>>>>>>>>>>>>>>>>>>>>>>>>>>>>>>>>>>>>>>>
\begin{picture}(65,20)(0,0) 
\put(0,-5){
%\put(0,0){\framebox(80,20){}}
%%
\put(0,5){                         
\begin{picture}(15,20)(0,0)
%%%%%%%%%%%%%%%%%%%%%%%%%%%%% potential + further left propagators
\multiput(15,3)(0,2){6}{$\cdot$}
\put(0,15){\line(1,0){5}}
\put(0,5){\vector(1,0){10}}
\put(15,15){\vector(-1,0){10}}
\put(15,5){\line(-1,0){5}}
\end{picture}}
\put(15,5){                         
\begin{picture}(25,20)(0,0)
%%%%%%%%%%%%%%%%%%%%%%%%%%%%% left propagators + box
\put(0,15){\line(1,0){5}}
\put(0,5){\vector(1,0){10}}
\put(15,15){\vector(-1,0){10}}
\put(15,5){\line(-1,0){5}}
\put(15,0){\framebox(10,20){}}
\end{picture}}
%%%%%%%%%%%%%%%%%%%%%%%%%% right propagators
\put(40,5){
\begin{picture}(15,20)(0,0)
\put(0,15){\line(1,0){5}}
\put(0,5){\vector(1,0){10}}
\put(15,15){\vector(-1,0){10}}
\put(15,5){\line(-1,0){5}}
\end{picture}}
}
\end{picture}
%<<<<<<<<<<<<<<<<<<<<<<<<<<<<<<<<<<<<<<<<<<<<<<<<<<<<<<<<<<<<<
%
\label{ladder}
\end{equation}
where the dotted line corresponds to the chiral
invariant quark-quark interaction of vertex $V$
and kernel $\cal K$. The arrowed line corresponds
to the Dirac quark propagator. The Ward identities
\cite{vector axial}
show that the ladder is consistent with the quark self 
energy equation in the rainbow approximation. 
When chiral symmetry is
spontaneously broken, the dressed 
quark propagator is non trivial,
\begin{equation} 
S = {i \over A(k^2) \not k - B(k^2) } \ ,
\end{equation} 
and the $\pi$ is a Goldstone boson in the chiral limit.
There is also evidence 
that the hadron-hadron 
coupled channel equations should include one meson
exchange, 
both in the sigma model 
\cite{Weinberg,miracle},
in the Nambu and Jona-Lasinio model 
\cite{Veronique}, 
in the constituent quark models
\cite{pi-pi},
and in an Euclidean quark model 
\cite{pi-pi}.
In microscopic calculations the Feynman diagram
for $\pi-\pi$ scattering must therefore 
include inside the box a vertical scalar ladder and a 
horizontal scalar ladder, see paper
\cite{pi-pi}
for details.
To cure double counting we also have to subtract
an empty box. 
\par
The main task of this paper
consists in computing 
{\em independently of the Quark Model},
and up to order $M_\pi^2$ and $P_i \, P_j$
in the $\pi$ mass and momenta,
the Feynman loop,
%
%LOOP
%
\begin{equation}
%>>>>>>>>>>>>>>>>>>>>>>>>>>>>>>>>>>>>>>>>>>>>>>
\begin{picture}(55,55)(0,0)
\put(0,-8){
%\put(0,0){\framebox(55,65){}}
%%%%%%%%%%%%%%%%%%%%%%%%%%%%%% box with two gammas
\put(16,13){\begin{picture}(20,40)(0,0)
\put(5,0){\line(0,1){5}}
\put(5,15){\vector(0,-1){10}}
\put(15,15){\line(0,-1){5}}
\put(15,0){\vector(0,1){10}}
\put(0,15){\framebox(20,10){}}
\put(5,25){\line(0,1){5}}
\put(5,40){\vector(0,-1){10}}
\put(15,40){\line(0,-1){5}}
\put(15,25){\vector(0,1){10}}
\end{picture}}
\put(16,8){\begin{picture}(20,40)(0,0)
\put(10,40){\oval(10,26)[t]}
\put(2,42){$\bullet$}
\put(-13,48){$\chi_{P_4}$}
\put(13,42){$\bullet$}
\put(18,48){$\chi_{P_1}$}
\put(10,53){\vector(-1,0){2}}
\end{picture}}
\put(16,-27){\begin{picture}(20,40)(0,0)
\put(10,45){\oval(10,26)[b]}
\put(2,38){$\bullet$}
\put(-15,34){$\chi_{P_3}$}
\put(13,38){$\bullet$}
\put(18,34){$\chi_{P_2}$}
\put(10,32){\vector(1,0){2}}
\end{picture}}
}
\end{picture}
%<<<<<<<<<<<<<<<<<<<<<<<<<<<<<<<<<<<<<<<<<<<<<<<<<<<
+
%>>>>>>>>>>>>>>>>>>>>>>>>>>>>>>>>>>>>>>>>>>>>>>
\begin{picture}(80,30)(0,0)
\put(0,-8){
%\put(0,0){\framebox(90,40){}}
%%%%%%%%%%%%%%%%%%%%%%%%%%%%%% box with two gammas
\put(20,10){\begin{picture}(40,20)(0,0)
\put(0,15){\line(1,0){5}}
\put(0,5){\vector(1,0){10}}
\put(15,15){\vector(-1,0){10}}
\put(15,5){\line(-1,0){5}}
\put(15,0){\framebox(10,20){}}
\put(25,15){\line(1,0){5}}
\put(25,5){\vector(1,0){10}}
\put(40,15){\vector(-1,0){10}}
\put(40,5){\line(-1,0){5}}
\end{picture}}
\put(20,10){\begin{picture}(40,20)(0,0)
\put(40,10){\oval(30,10)[r]}
\put(42,2){$\bullet$}
\put(38,-5){$\chi_{P_2}$}
\put(42,13){$\bullet$}
\put(38,21){$\chi_{P_1}$}
\put(55,10){\vector(0,1){2}}
\end{picture}}
\put(-25,10){\begin{picture}(40,20)(0,0)
\put(45,10){\oval(30,10)[l]}
\put(38,2){$\bullet$}
\put(34,-5){$\chi_{P_3}$}
\put(38,13){$\bullet$}
\put(34,21){$\chi_{P_4}$}
\put(30,10){\vector(0,-1){2}}
\end{picture}}
}
\end{picture}
%<<<<<<<<<<<<<<<<<<<<<<<<<<<<<<<<<<<<<<<<<<<<<<<<<<<
-
%>>>>>>>>>>>>>>>>>>>>>>>>>>>>>>>>>>>>>>>>>>>>>>>>>>>
\begin{picture}(55,30)(0,0)
\put(0,-8){
%\put(0,0){\framebox(65,40){}}
%%%%%%%%%%%%%%%%%%%%%%%%%%%%%% four Gammas
\put(20,10){\begin{picture}(40,20)(0,0)
\put(0,15){\line(1,0){5}}
\put(0,5){\vector(1,0){10}}
\put(15,15){\vector(-1,0){10}}
\put(15,5){\line(-1,0){5}}
\end{picture}}
\put(-5,10){\begin{picture}(40,20)(0,0)
\put(40,10){\oval(30,10)[r]}
\put(42,2){$\bullet$}
\put(38,-5){$\chi_{P_2}$}
\put(42,13){$\bullet$}
\put(38,21){$\chi_{P_1}$}
\put(55,10){\vector(0,1){2}}
\end{picture}}
\put(-25,10){\begin{picture}(40,20)(0,0)
\put(45,10){\oval(30,10)[l]}
\put(38,2){$\bullet$}
\put(34,-5){$\chi_{P_3}$}
\put(38,13){$\bullet$}
\put(34,21){$\chi_{P_4}$}
\put(30,10){\vector(0,-1){2}}
\end{picture}}
}
\end{picture}
%<<<<<<<<<<<<<<<<<<<<<<<<<<<<<<<<<<<<<<<<<<<<<<<<<<<
\ ,
\label{loop}
\end{equation}
where $\chi$ is
the Bethe Salpeter vertex of the pion. The
subindex $_{Pi}$ accounts an external momentum flowing into
the loop.
\par
To derive the proof it is convenient to define the
dressed axial vertex $\Gamma_{\hspace{-.08cm}A}$
with the axial Ward identity,
\begin{equation}
\Gamma_{\hspace{-.08cm}A}(k_1,k_2) = S^{-1}(k_1) \gamma_5 + \gamma_5 S^{-1}(k_2) \ ,
\end{equation}
because it is related to the pion vertex and
because it has the crucial property,
%
%CRUCIAL
%
\begin{eqnarray}
\label{crucial}
%>>>>>>>>>>>>>>>>>>>>>>>>>>>>>>>>>>>
\begin{picture}(86,30)(0,0)
\put(0,-8){
%\put(0,0){\framebox(86,40){}}
%
%%%%%%%%%%%%%%%%%%%%%%%%%%%%%% box
\put(3,10){\begin{picture}(40,20)(0,0)
\put(0,15){\line(1,0){5}}
\put(0,5){\vector(1,0){10}}
\put(15,15){\vector(-1,0){10}}
\put(15,5){\line(-1,0){5}}
\put(15,0){\framebox(10,20){}}
\put(25,15){\line(1,0){5}}
\put(25,5){\vector(1,0){10}}
\put(40,15){\vector(-1,0){10}}
\put(40,5){\line(-1,0){5}}
\end{picture}}
%
%%%%%%%%%%%%%%%%% intermediate Gamma_{\hspace{-.08cm}A}
\put(3,10){\begin{picture}(60,20)(0,0)
\put(38,2){$\bullet$}
\put(35,-8){$S^{-1}$}
\put(38,13){$\bullet$}
\put(37,20){$\Gamma_{\hspace{-.08cm}A}$}
\end{picture}}
%
%%%%%%%%%%%%%%%%%%%%%%%%%%%%%% box
\put(43,10){\begin{picture}(40,20)(0,0)
\put(0,15){\line(1,0){5}}
\put(0,5){\vector(1,0){10}}
\put(15,15){\vector(-1,0){10}}
\put(15,5){\line(-1,0){5}}
\put(15,0){\framebox(10,20){}}
\put(25,15){\line(1,0){5}}
\put(25,5){\vector(1,0){10}}
\put(40,15){\vector(-1,0){10}}
\put(40,5){\line(-1,0){5}}
\end{picture}}
}
\end{picture}
%<<<<<<<<<<<<<<<<<<<<<<<<<<<<<<<<<<<
&=&
%>>>>>>>>>>>>>>>>>>>>>>>>>>>>>>>>>>>>>>>>>
\begin{picture}(56,30)(0,0)
\put(0,-8){
%\put(0,0){\framebox(56,30){}}
%%%%%%%%%%%%%%%%%%%%%%%%%%%%%% box
\put(13,5){\begin{picture}(40,20)(0,0)
\put(0,15){\line(1,0){5}}
\put(0,5){\vector(1,0){10}}
\put(15,15){\vector(-1,0){10}}
\put(15,5){\line(-1,0){5}}
\put(15,0){\framebox(10,20){}}
\put(25,15){\line(1,0){5}}
\put(25,5){\vector(1,0){10}}
\put(40,15){\vector(-1,0){10}}
\put(40,5){\line(-1,0){5}}
\put(-11,15){$\gamma_5 $}
\end{picture}}
}
\end{picture}
%<<<<<<<<<<<<<<<<<<<<<<<<<<<<<<<<
+
%>>>>>>>>>>>>>>>>>>>>>>>>>>>>>>>>>>>>>>>
\begin{picture}(54,30)(0,0)
\put(0,-8){
%\put(0,0){\framebox(54,30){}}
%%%%%%%%%%%%%%%%%%%%%%%%%%%%%% box
\put(3,5){\begin{picture}(40,20)(0,0)
\put(0,15){\line(1,0){5}}
\put(0,5){\vector(1,0){10}}
\put(15,15){\vector(-1,0){10}}
\put(15,5){\line(-1,0){5}}
\put(15,0){\framebox(10,20){}}
\put(25,15){\line(1,0){5}}
\put(25,5){\vector(1,0){10}}
\put(40,15){\vector(-1,0){10}}
\put(40,5){\line(-1,0){5}}
\put(40,15){$\gamma_5 $}
\end{picture}}
}
\end{picture}
%<<<<<<<<<<<<<<<<<<<<<<<<<<<<<<<<<<<<<<
\end{eqnarray}
which constitutes a Ward identity for the ladder.
This identity is derived if we expand the ladders and 
substitute the vertex in the left hand side. Then all terms with 
an intermediate $\gamma_5$ include the anticommutator $\{\gamma_5,V\}$
and this cancels because {\em the interaction is chiral
invariant} and {\em the kernel is local}. 
Only the right hand side survives.

\par
We now derive a second useful relation for $\Gamma_A$.
The Dyson Schwinger equation 
for the full propagator is,
\begin{equation}
S(k)^{-1}=S_0^{-1}(k)-\int {d^4 q \over (2 \pi)^4}
{\cal K}(q) V S(k+q) V   
\label{SD}
\end{equation}
where $V$ is some vertex that anticommutes with
$\gamma_5$ and that is color dependent as well, 
$\cal K$ is the quark-quark Kernel, and 
 $S_0= i / ( \not k + m )$ is the bare quark propagator.
If we multiply right or left with $\gamma_5$
and sum, we get,  
\begin{eqnarray}
& S(k_1)^{-1}\gamma_5 + \gamma_5 S(k_2)^{-1} 
= S_0(k_1)^{-1}\gamma_5 + \gamma_5 S_0(k_2)^{-1} &
\nonumber \\  
& + \int {\cal K}(q) V ( S(k_1+q)\gamma_5 + \gamma_5 S(k_2+q) ) V \ , & 
\end{eqnarray}
and this is the Bethe Salpeter equation for the vertex,
\FL
\begin{eqnarray}
\label{BS vertex}
\Gamma_{\hspace{-.08cm}A}(k_1,k_2) &=& \hspace{-.08cm} 
\gamma_{\hspace{-.08cm}A}(k_1,k_2) 
+\hspace{-.08cm} \int {\cal K}(q) V S(k_1+q)\Gamma_{\hspace{-.08cm}A} 
S(k_2+q) V 
\nonumber \\
\gamma_{\hspace{-.08cm}A}(k_1,k_2)&=&  S^{-1}_0(k_1) \gamma_5 
+ \gamma_5 S^{-1}_0(k_2) ,
\end{eqnarray}
where the particular part $\gamma_{\hspace{-.08cm}A}$ is
the bare axial vertex,
\begin{equation}
\gamma_{\hspace{-.08cm}A} = 
{ ( \not P  - 2 m )\over i} \gamma_5 \ , \ \
P=k_1-k_2 \ .
\label{bare}
\end{equation} 
$\gamma_{\hspace{-.08cm}A}$
vanishes when the current quark mass $m$ is small (chiral limit) and 
at the same time the total momentum $P^\mu$ of the vertex is small.
On the other hand the dressed vertex $\Gamma_{\hspace{-.08cm}A}$ is finite,
\FL
\begin{equation}
\Gamma_{\hspace{-.08cm}A}(k_1,k_2) \hspace{-.08cm} = \hspace{-.08cm} 
{  A(k_1) \hspace{-.1cm} 
\not k_1 \hspace{-.05cm} - \hspace{-.05cm} A(k_2) \hspace{-.1cm} \not k_2 \hspace{-.05cm}
- \hspace{-.05cm} B(k_1) \hspace{-.05cm} - \hspace{-.05cm} B(k_2) 
\over i} \gamma_5 , \hspace{-.3cm}
\end{equation}
providing spontaneous chiral symmetry breaking occurs in
eq. (\ref{SD})
to generate a dynamical mass in the dressed quark propagator.
After we iterate the Bethe Salpeter equation for the
dressed axial vertex $\Gamma_{\hspace{-.08cm}A}$, and include the external
propagators, we get the second useful relation,
%
%SECOND
%
\begin{equation}
S \, \Gamma_{\hspace{-.08cm}A} \, S
= 
%>>>>>>>>>>>>>>>>>>>>>>>>>>>>>>>>>>>>>>>>>>>>>>>>>>
\begin{picture}(70,20)(0,0)
\put(0,-8){
%\put(0,0){\framebox(70,25){}}
\put(5,10){
%%%%%%%%%%%%%%%%%%%%%%%%%%%%%% half closed box
\begin{picture}(40,20)(0,7)
\put(0,15){\line(1,0){5}}
\put(0,5){\vector(1,0){10}}
\put(15,15){\vector(-1,0){10}}
\put(15,5){\line(-1,0){5}}
\put(15,0){\framebox(10,20){}}
\put(25,15){\line(1,0){5}}
\put(25,5){\vector(1,0){10}}
\put(40,15){\vector(-1,0){10}}
\put(40,5){\line(-1,0){5}}
\put(40,10){\oval(10,10)[r]}
\put(43,8){$\bullet$}
\put(49,10){$\gamma_{\hspace{-.08cm}A}$}
\end{picture}
}
}
\end{picture}
%<<<<<<<<<<<<<<<<<<<<<<<<<<<<<<<<<<<<<<<<<<<<<
\ .
\label{second}
\end{equation}

\par
The main step to get the proof consists in 
decreasing the number of vertices with the 
help of eqs. (\ref{crucial}) and (\ref{second}),
%
%STEP
%
\begin{eqnarray}
%>>>>>>>>>>>>>>>>>>>>>>>>>>>>>>>>>>>>>>>>>>>>>>
\begin{picture}(70,30)(0,0)
\put(0,-8){
%\put(0,0){\framebox(70,40){}}
\put(5,10){
%%%%%%%%%%%%%%%%%%%%%%%%%%%%%% box with two gamma_{\hspace{-.08cm}A}s
\begin{picture}(40,20)(0,0)
\put(0,15){\line(1,0){5}}
\put(0,5){\vector(1,0){10}}
\put(15,15){\vector(-1,0){10}}
\put(15,5){\line(-1,0){5}}
\put(15,0){\framebox(10,20){}}
\put(25,15){\line(1,0){5}}
\put(25,5){\vector(1,0){10}}
\put(40,15){\vector(-1,0){10}}
\put(40,5){\line(-1,0){5}}
\put(40,10){\oval(30,10)[r]}
\put(42,2){$\bullet$}
\put(48,-6){$\Gamma_{\hspace{-.08cm}A}$}
\put(42,13){$\bullet$}
\put(48,18){$\Gamma_{\hspace{-.08cm}A}$}
\put(55,10){\vector(0,1){2}}
\end{picture}
}
}
\end{picture}
%<<<<<<<<<<<<<<<<<<<<<<<<<<<<<<<<<<<<<<<<<<<<<<<<<<<
&=&
%>>>>>>>>>>>>>>>>>>>>>>>>>>>>>>>>>>>>>>>>>>>>>>>>>
\begin{picture}(105,30)(0,0)
\put(0,-8){
%\put(0,0){\framebox(105,40){}}
%%%%%%%%%%%%%%%%%%%%%%%%%%%%%% box
\put(5,10){\begin{picture}(40,20)(0,0)
\put(0,15){\line(1,0){5}}
\put(0,5){\vector(1,0){10}}
\put(15,15){\vector(-1,0){10}}
\put(15,5){\line(-1,0){5}}
\put(15,0){\framebox(10,20){}}
\put(25,15){\line(1,0){5}}
\put(25,5){\vector(1,0){10}}
\put(40,15){\vector(-1,0){10}}
\put(40,5){\line(-1,0){5}}
\end{picture}}
%%%%%%%%%%%%%%%%% intermediate Gamma_{\hspace{-.08cm}A}
\put(5,10){\begin{picture}(60,20)(0,0)
\put(38,2){$\bullet$}
\put(35,-8){$S^{-1}$}
\put(38,13){$\bullet$}
\put(37,20){$\Gamma_{\hspace{-.08cm}A}$}
\end{picture}}
%
%%%%%%%%%%%%%%%%%%%%%%%%%%%%%% box
\put(45,10){\begin{picture}(40,20)(0,0)
\put(0,15){\line(1,0){5}}
\put(0,5){\vector(1,0){10}}
\put(15,15){\vector(-1,0){10}}
\put(15,5){\line(-1,0){5}}
\put(15,0){\framebox(10,20){}}
\put(25,15){\line(1,0){5}}
\put(25,5){\vector(1,0){10}}
\put(40,15){\vector(-1,0){10}}
\put(40,5){\line(-1,0){5}}
\end{picture}}
%%%%%%%%%%%%%%%%%%%%%%%%%%%%%% half closed
\put(45,10){\begin{picture}(40,20)(0,0)
\put(40,10){\oval(10,10)[r]}
\put(43,8){$\bullet$}
\put(49,10){$\gamma_{\hspace{-.08cm}A}$}
\end{picture}}
}
\end{picture}
%<<<<<<<<<<<<<<<<<<<<<<<<<<<<<<<<<<<<<<<<<<<<<<<
\label{step}
\\ \nonumber
&=&
%>>>>>>>>>>>>>>>>>>>>>>>>>>>>>>>>>>>>>>>>>>>>>>>>>
\begin{picture}(55,20)(0,0)
\put(0,-13){
%\put(0,0){\framebox(55,30){}}
%%%%%%%%%%%%%%%%%%%%%%%%%%%%%% just Gamma_{\hspace{-.08cm}A}
\put(-5,5){\begin{picture}(40,20)(0,0)
\put(25,15){\line(1,0){5}}
\put(25,5){\vector(1,0){10}}
\put(40,15){\vector(-1,0){10}}
\put(40,5){\line(-1,0){5}}
\end{picture}}
%%%%%%%%%%%%%%%%%%%%%%%%%%%%%% half closed
\put(-5,5){\begin{picture}(40,20)(0,0)
\put(40,10){\oval(10,10)[r]}
\put(43,8){$\bullet$}
\put(10,15){$\gamma_5$}
\put(49,10){$\Gamma_{\hspace{-.08cm}A}$}
\end{picture}}
}
\end{picture}
%<<<<<<<<<<<<<<<<<<<<<<<<<<<<<<<<<<<<<<<<<
+
%>>>>>>>>>>>>>>>>>>>>>>>>>>>>>>>>>>>>>>>>
\begin{picture}(75,20)(0,0)
\put(0,-13){
%\put(0,0){\framebox(75,30){}}
%%%%%%%%%%%%%%%%%%%%%%%%%%%%%% box
\put(5,5){\begin{picture}(40,20)(0,0)
\put(0,15){\line(1,0){5}}
\put(0,5){\vector(1,0){10}}
\put(15,15){\vector(-1,0){10}}
\put(15,5){\line(-1,0){5}}
\put(15,0){\framebox(10,20){}}
\put(25,15){\line(1,0){5}}
\put(25,5){\vector(1,0){10}}
\put(40,15){\vector(-1,0){10}}
\put(40,5){\line(-1,0){5}}
\end{picture}}
%%%%%%%%%%%%%%%%%%%%%%%%%%%%%% half closed
\put(5,5){\begin{picture}(40,20)(0,0)
\put(40,10){\oval(10,10)[r]}
\put(43,8){$\bullet$}
\put(49,10){$\gamma_5 \gamma_{\hspace{-.08cm}A}$}
\end{picture}}
}
\end{picture}
%<<<<<<<<<<<<<<<<<<<<<<<<<<<<<<<<<<<<<<<<<<<<<<<<<<<<
\ .
\end{eqnarray}
This is repeated to compute the square box with ladder,
%
%BOX WITH LADDER
%
\begin{eqnarray}
%>>>>>>>>>>>>>>>>>>>>>>>>>>>>>>>>>>>>>>>>>>>>>>
\begin{picture}(80,30)(0,0)
\put(0,-8){
%\put(0,0){\framebox(90,40){}}
%%%%%%%%%%%%%%%%%%%%%%%%%%%%%% box with two gammas
\put(20,10){\begin{picture}(40,20)(0,0)
\put(0,15){\line(1,0){5}}
\put(0,5){\vector(1,0){10}}
\put(15,15){\vector(-1,0){10}}
\put(15,5){\line(-1,0){5}}
\put(15,0){\framebox(10,20){}}
\put(25,15){\line(1,0){5}}
\put(25,5){\vector(1,0){10}}
\put(40,15){\vector(-1,0){10}}
\put(40,5){\line(-1,0){5}}
\end{picture}}
\put(20,10){\begin{picture}(40,20)(0,0)
\put(40,10){\oval(30,10)[r]}
\put(42,2){$\bullet$}
\put(38,-5){${\Gamma_{\hspace{-.08cm}A}}_{P_2}$}
\put(42,13){$\bullet$}
\put(38,21){${\Gamma_{\hspace{-.08cm}A}}_{P_1}$}
\put(55,10){\vector(0,1){2}}
\end{picture}}
\put(-25,10){\begin{picture}(40,20)(0,0)
\put(45,10){\oval(30,10)[l]}
\put(38,2){$\bullet$}
\put(34,-5){${\Gamma_{\hspace{-.08cm}A}}_{P_3}$}
\put(38,13){$\bullet$}
\put(34,21){${\Gamma_{\hspace{-.08cm}A}}_{P_4}$}
\put(30,10){\vector(0,-1){2}}
\end{picture}}
}
\end{picture}
%<<<<<<<<<<<<<<<<<<<<<<<<<<<<<<<<<<<<<<<<<<<<<<<<<<<
&=&
{1 \over 2}
%>>>>>>>>>>>>>>>>>>>>>>>>>>>>>>>>>>>>>>>>>>>>>>>>>>>
\begin{picture}(65,30)(0,0)
\put(0,-8){
%\put(0,0){\framebox(65,40){}}
%%%%%%%%%%%%%%%%%%%%%%%%%%%%%% four Gamma_{\hspace{-.08cm}A}s
\put(25,10){\begin{picture}(40,20)(0,0)
\put(0,15){\line(1,0){5}}
\put(0,5){\vector(1,0){10}}
\put(15,15){\vector(-1,0){10}}
\put(15,5){\line(-1,0){5}}
\end{picture}}
\put(0,10){\begin{picture}(40,20)(0,0)
\put(40,10){\oval(30,10)[r]}
\put(42,2){$\bullet$}
\put(48,-6){$\Gamma_{\hspace{-.08cm}A}$}
\put(42,13){$\bullet$}
\put(48,18){$\Gamma_{\hspace{-.08cm}A}$}
\put(55,10){\vector(0,1){2}}
\end{picture}}
\put(-20,10){\begin{picture}(40,20)(0,0)
\put(45,10){\oval(30,10)[l]}
\put(38,2){$\bullet$}
\put(34,-6){$\Gamma_{\hspace{-.08cm}A}$}
\put(38,13){$\bullet$}
\put(34,18){$\Gamma_{\hspace{-.08cm}A}$}
\put(30,10){\vector(0,-1){2}}
\end{picture}}
}
\end{picture}
%<<<<<<<<<<<<<<<<<<<<<<<<<<<<<<<<<<<<<<<<<<<<<<<<<<<
\label{box with ladder}
\\
+
2 
%>>>>>>>>>>>>>>>>>>>>>>>>>>>>>>>>>>>>>>>>>>>>>>>>>>>
\begin{picture}(60,15)(0,0)
\put(0,-18){
%\put(0,10){\framebox(60,20){}}
%%%%%%%%%%%%%%%%%%%%%%%%%%%%%% loop 
\put(5,10){\begin{picture}(40,20)(0,0)
\put(40,10){\oval(12,12)}
\put(46,10){\vector(0,1){2}}
\put(32,8){$\bullet$}
\put(-2,10){$\{ \gamma_{\hspace{-.08cm}A} , \gamma_5 \} $}
\end{picture}}
}
\end{picture}
%<<<<<<<<<<<<<<<<<<<<<<<<<<<<<<<<<<<<<<<<<<<<<<<<<<<
&+&
{1 \over 4} 
%>>>>>>>>>>>>>>>>>>>>>>>>>>>>>>>>>>>>>>>>>>>>>>>>>>>
\begin{picture}(125,15)(0,0)
\put(0,-18){
%\put(0,0){\framebox(125,40){}}
%%%%%%%%%%%%%%%%%%%%%%%%%%%%%% box with two gamma_{\hspace{-.08cm}A}s
\put(40,10){\begin{picture}(40,20)(0,0)
\put(0,15){\line(1,0){5}}
\put(0,5){\vector(1,0){10}}
\put(15,15){\vector(-1,0){10}}
\put(15,5){\line(-1,0){5}}
\put(15,0){\framebox(10,20){}}
\put(25,15){\line(1,0){5}}
\put(25,5){\vector(1,0){10}}
\put(40,15){\vector(-1,0){10}}
\put(40,5){\line(-1,0){5}}
\end{picture}}
%
%%%%%%%%%%%%%%%%%%%%%%%%%%%%%% half closed
\put(40,10){\begin{picture}(40,20)(0,0)
\put(40,10){\oval(10,10)[r]}
\put(43,8){$\bullet$}
\put(49,10){$\{ \gamma_{\hspace{-.08cm}A} , \gamma_5 \} $}
\end{picture}}
\put(-5,10){\begin{picture}(40,20)(0,0)
\put(45,10){\oval(10,10)[l]}
\put(38,8){$\bullet$}
\put(5,10){$\{ \gamma_{\hspace{-.08cm}A} , \gamma_5 \} $}
\end{picture}}
}
\end{picture}
%<<<<<<<<<<<<<<<<<<<<<<<<<<<<<<<<<<<<<<<<<<<<<<<<<<<
\ ,
\nonumber
\end{eqnarray}
and we get three terms, respectively of order 
$1$, $\gamma_{\hspace{-.08cm}A}$ and $\gamma_{\hspace{-.08cm}A}^2$.
We note that all the other factors 
( $\Gamma_{\hspace{-.08cm}A}$, $S$, and the scalar and vector ladder) 
are finite and carry the scale of the effective quark-quark
interaction. 

\par
When we sum the three diagrams of the full box,
\begin{eqnarray}
%>>>>>>>>>>>>>>>>>>>>>>>>>>>>>>>>>>>>>>>>>>>>>>
\begin{picture}(55,55)(0,8)
\put(0,0){
%\put(0,0){\framebox(55,65){}}
%%%%%%%%%%%%%%%%%%%%%%%%%%%%%% box with two gammas
\put(16,13){\begin{picture}(20,40)(0,0)
\put(5,0){\line(0,1){5}}
\put(5,15){\vector(0,-1){10}}
\put(15,15){\line(0,-1){5}}
\put(15,0){\vector(0,1){10}}
\put(0,15){\framebox(20,10){}}
\put(5,25){\line(0,1){5}}
\put(5,40){\vector(0,-1){10}}
\put(15,40){\line(0,-1){5}}
\put(15,25){\vector(0,1){10}}
\end{picture}}
\put(16,8){\begin{picture}(20,40)(0,0)
\put(10,40){\oval(10,26)[t]}
\put(2,40){$\bullet$}
\put(-15,49){${\Gamma_{\hspace{-.08cm}A}}_{P_4}$}
\put(13,42){$\bullet$}
\put(18,48){${\Gamma_{\hspace{-.08cm}A}}_{P_1}$}
\put(10,53){\vector(-1,0){2}}
\end{picture}}
\put(16,-27){\begin{picture}(20,40)(0,0)
\put(10,45){\oval(10,26)[b]}
\put(2,38){$\bullet$}
\put(-15,34){${\Gamma_{\hspace{-.08cm}A}}_{P_3}$}
\put(13,38){$\bullet$}
\put(18,34){${\Gamma_{\hspace{-.08cm}A}}_{P_2}$}
\put(10,32){\vector(1,0){2}}
\end{picture}}
}
\end{picture}
%<<<<<<<<<<<<<<<<<<<<<<<<<<<<<<<<<<<<<<<<<<<<<<<<<<<
+
%>>>>>>>>>>>>>>>>>>>>>>>>>>>>>>>>>>>>>>>>>>>>>>
\begin{picture}(80,30)(0,0)
\put(0,0){
%\put(0,0){\framebox(90,40){}}
%%%%%%%%%%%%%%%%%%%%%%%%%%%%%% box with two gammas
\put(20,10){\begin{picture}(40,20)(0,0)
\put(0,15){\line(1,0){5}}
\put(0,5){\vector(1,0){10}}
\put(15,15){\vector(-1,0){10}}
\put(15,5){\line(-1,0){5}}
\put(15,0){\framebox(10,20){}}
\put(25,15){\line(1,0){5}}
\put(25,5){\vector(1,0){10}}
\put(40,15){\vector(-1,0){10}}
\put(40,5){\line(-1,0){5}}
\end{picture}}
\put(20,10){\begin{picture}(40,20)(0,0)
\put(40,10){\oval(30,10)[r]}
\put(42,2){$\bullet$}
\put(38,-5){${\Gamma_{\hspace{-.08cm}A}}_{P_2}$}
\put(42,13){$\bullet$}
\put(38,21){${\Gamma_{\hspace{-.08cm}A}}_{P_1}$}
\put(55,10){\vector(0,1){2}}
\end{picture}}
\put(-25,10){\begin{picture}(40,20)(0,0)
\put(45,10){\oval(30,10)[l]}
\put(38,2){$\bullet$}
\put(34,-5){${\Gamma_{\hspace{-.08cm}A}}_{P_3}$}
\put(38,13){$\bullet$}
\put(34,21){${\Gamma_{\hspace{-.08cm}A}}_{P_4}$}
\put(30,10){\vector(0,-1){2}}
\end{picture}}
}
\end{picture}
%<<<<<<<<<<<<<<<<<<<<<<<<<<<<<<<<<<<<<<<<<<<<<<<<<<<
-
%>>>>>>>>>>>>>>>>>>>>>>>>>>>>>>>>>>>>>>>>>>>>>>>>>>>
\begin{picture}(55,30)(0,0)
\put(0,0){
%\put(0,0){\framebox(65,40){}}
%%%%%%%%%%%%%%%%%%%%%%%%%%%%%% four Gammas
\put(20,10){\begin{picture}(40,20)(0,0)
\put(0,15){\line(1,0){5}}
\put(0,5){\vector(1,0){10}}
\put(15,15){\vector(-1,0){10}}
\put(15,5){\line(-1,0){5}}
\end{picture}}
\put(-5,10){\begin{picture}(40,20)(0,0)
\put(40,10){\oval(30,10)[r]}
\put(42,2){$\bullet$}
\put(38,-5){${\Gamma_{\hspace{-.08cm}A}}_{P_2}$}
\put(42,13){$\bullet$}
\put(38,21){${\Gamma_{\hspace{-.08cm}A}}_{P_1}$}
\put(55,10){\vector(0,1){2}}
\end{picture}}
\put(-25,10){\begin{picture}(40,20)(0,0)
\put(45,10){\oval(30,10)[l]}
\put(38,2){$\bullet$}
\put(34,-5){${\Gamma_{\hspace{-.08cm}A}}_{P_3}$}
\put(38,13){$\bullet$}
\put(34,21){${\Gamma_{\hspace{-.08cm}A}}_{P_4}$}
\put(30,10){\vector(0,-1){2}}
\end{picture}}
}
\end{picture}
%<<<<<<<<<<<<<<<<<<<<<<<<<<<<<<<<<<<<<<<<<<<<<<<<<<<
\label{ciclic 1}
\\
= 8 \, i \, m \, tr \{ S \} + o(m^2,P_i^4,mP_i^2) \ ,
\label{high Ward}
\end{eqnarray}
the finite term of order $1$ {\em exactly cancels}. 
The total box vanishes in the chiral
limit of $m=0$, although each one of the three boxes is
generally finite and proportional to the fourth power of 
the scale of the interaction. This cancellation  
in eq. (\ref{high Ward}) complies with the Adler zero
and corresponds to a higher Ward identity. 

\par
The dependence of eq. (\ref{high Ward}) on the quark current mass $m$
is trivially computed with eqs. (\ref{bare}) and (\ref{box with ladder}). 
The exact cancellation in the momentum dependence is more subtle.
Expanding up the first order in $P_i P_j$, the  $\gamma_{\hspace{-.08cm}A}$ 
and  $\gamma_{\hspace{-.08cm}A}^2$ terms in eq. (\ref{box with ladder}),
\begin{equation}
\nonumber \\
%
%>>>>>>>>>>>>>>>>>>>>>>>>>>>>>>>>>>>>>>>>>>>>>>>>>>>
\begin{picture}(217,35)(0,8)
\put(0,0){
%\put(0,0){\framebox(217,40){}}
%%%%%%%%%%%%%%%%%%%%%%%%%%%%%% box with two gamma_{\hspace{-.08cm}A}s
\put(130,10){\begin{picture}(40,20)(0,0)
\put(0,15){\line(1,0){5}}
\put(0,5){\vector(1,0){10}}
\put(15,15){\vector(-1,0){10}}
\put(15,5){\line(-1,0){5}}
\put(15,0){\framebox(10,20){}}
\put(25,15){\line(1,0){5}}
\put(25,5){\vector(1,0){10}}
\put(40,15){\vector(-1,0){10}}
\put(40,5){\line(-1,0){5}}
\end{picture}}
%
%%%%%%%%%%%%%%%%%%%%%%%%%%%%%% 1st half open with propagators
\put(83,10){\begin{picture}(40,20)(0,0)
\put(0,15){\line(1,0){5}}
\put(0,5){\vector(1,0){10}}
\put(15,15){\vector(-1,0){10}}
\put(15,5){\line(-1,0){5}}
\put(0,10){\oval(10,10)[l]}
\put(-8,8){$\bullet$}
\put(-45,10){$-i \partial_\mu S^{-1} $}
\end{picture}}
%
%%%%%%%%%%%%%%%%%%%%%%%%%%%%%% half closed
\put(85,10){\begin{picture}(40,20)(0,0)
\put(45,10){\oval(10,10)[l]}
\put(37,8){$\bullet$}
\put(15,10){$+ \,\gamma_\mu$}
\end{picture}}
%
%%%%%%%%%%%%%%%%%%%%%%%%%%%%%% half closed
\put(0,10){\begin{picture}(40,20)(0,0)
\put(0,10){${P_3^\mu-P_4^\mu \over i}$}
\put(32,10){$\Biggl[$}
\end{picture}}
%
%%%%%%%%%%%%%%%%%%%%%%%%%%%%% right side
\put(170,10){\begin{picture}(40,20)(0,0)
\put(10,10){\oval(10,10)[r]}
\put(13,8){$\bullet$}
\put(19,10){${\not P_1 - \not P_2 \over i} $}
\put(3,10){$\Biggr]$}
\end{picture}}
}
\end{picture}
%<<<<<<<<<<<<<<<<<<<<<<<<<<<<<<<<<<<<<<<<<<<<<<<<<<<
\end{equation} 
and this cancels due to the vector Ward Identity which
has and analogous variant to eq. (\ref{second}).

\par
We now apply these Ward identity techniques
to the actual Feynman amplitude with pion vertices of 
eq. (\ref{loop}). 
In the limit of vanishing current quark mass $m$ and 
vertex momentum $P^\mu$, the Bethe-Salpeter (\ref{BS vertex})
equation for the 
axial vertex $\Gamma_A$ becomes homegenous and is
thus identical to the homogeneous 
Bethe-Salpeter equation for the
pion vertex ${\chi_\pi}_P(k)$. In this limit the pion is
a massless Goldstone boson, and the pion Bethe Salpeter 
vertex is proportional to the dressed axial vertex,
and to the dynamical quark mass $B(k)$,
\begin{equation}
\chi_{_0}(k)= { B(k) \over n_\pi }\gamma_5 
= {1 \over 2 \, i \, n_\pi } {\Gamma_{\hspace{-.08cm}A}}_0(k,k) \ ,
\end{equation}
where $n_\pi$ is the norm. And the loop (\ref{loop}) {\em vanishes}.
\par
To get the loop (\ref{loop}) up to order $P^2_i$ and
$M_\pi^2$, we need
at most two full Bethe Salpeter vertices
$\chi$, the other two can be approximated by 
$\Gamma_{\hspace{-.08cm}A} / (2 \, i \, n_\pi)$. 
Expanding up to second order in 
$ \chi-{ \Gamma_{\hspace{-.08cm}A} / (2 \, i \, n_\pi} ) $
and regrouping the sum, we find that the amplitude
of eq.(\ref{loop}) is the sum of four classes
of terms.  Each class includes a sum of the possible 
cyclic permutations of the external momenta 
$P_1, \, P_2, \, P_3,$ and $P_4$.
We get, eq.( \ref{ciclic 1} ) with factor 3 times $(2 \, i \, n_\pi)^{-4}$ ,
minus , with factor $2$ times $(2 \, i \, n_\pi)^{-3}$ , (4 permutations)
%
%CYCLIC 2
%
\begin{equation}
%>>>>>>>>>>>>>>>>>>>>>>>>>>>>>>>>>>>>>>>>>>>>>>
\begin{picture}(55,55)(0,0)
\put(0,-8){
%\put(0,0){\framebox(55,65){}}
%%%%%%%%%%%%%%%%%%%%%%%%%%%%%% box with two gammas
\put(16,13){\begin{picture}(20,40)(0,0)
\put(5,0){\line(0,1){5}}
\put(5,15){\vector(0,-1){10}}
\put(15,15){\line(0,-1){5}}
\put(15,0){\vector(0,1){10}}
\put(0,15){\framebox(20,10){}}
\put(5,25){\line(0,1){5}}
\put(5,40){\vector(0,-1){10}}
\put(15,40){\line(0,-1){5}}
\put(15,25){\vector(0,1){10}}
\end{picture}}
\put(16,8){\begin{picture}(20,40)(0,0)
\put(10,40){\oval(10,26)[t]}
\put(2,42){$\bullet$}
\put(-13,48){$\chi_{P_4}$}
\put(13,42){$\bullet$}
\put(18,48){${\Gamma_{\hspace{-.08cm}A}}_{P_1}$}
\put(10,53){\vector(-1,0){2}}
\end{picture}}
\put(16,-27){\begin{picture}(20,40)(0,0)
\put(10,45){\oval(10,26)[b]}
\put(2,38){$\bullet$}
\put(-15,34){${\Gamma_{\hspace{-.08cm}A}}_{P_3}$}
\put(13,38){$\bullet$}
\put(18,34){${\Gamma_{\hspace{-.08cm}A}}_{P_2}$}
\put(10,32){\vector(1,0){2}}
\end{picture}}
}
\end{picture}
%<<<<<<<<<<<<<<<<<<<<<<<<<<<<<<<<<<<<<<<<<<<<<<<<<<<
+
%>>>>>>>>>>>>>>>>>>>>>>>>>>>>>>>>>>>>>>>>>>>>>>
\begin{picture}(80,30)(0,0)
\put(0,-8){
%\put(0,0){\framebox(90,40){}}
%%%%%%%%%%%%%%%%%%%%%%%%%%%%%% box with two gammas
\put(20,10){\begin{picture}(40,20)(0,0)
\put(0,15){\line(1,0){5}}
\put(0,5){\vector(1,0){10}}
\put(15,15){\vector(-1,0){10}}
\put(15,5){\line(-1,0){5}}
\put(15,0){\framebox(10,20){}}
\put(25,15){\line(1,0){5}}
\put(25,5){\vector(1,0){10}}
\put(40,15){\vector(-1,0){10}}
\put(40,5){\line(-1,0){5}}
\end{picture}}
\put(20,10){\begin{picture}(40,20)(0,0)
\put(40,10){\oval(30,10)[r]}
\put(42,2){$\bullet$}
\put(38,-5){${\Gamma_{\hspace{-.08cm}A}}_{P_2}$}
\put(42,13){$\bullet$}
\put(38,21){${\Gamma_{\hspace{-.08cm}A}}_{P_1}$}
\put(55,10){\vector(0,1){2}}
\end{picture}}
\put(-25,10){\begin{picture}(40,20)(0,0)
\put(45,10){\oval(30,10)[l]}
\put(38,2){$\bullet$}
\put(34,-5){${\Gamma_{\hspace{-.08cm}A}}_{P_3}$}
\put(38,13){$\bullet$}
\put(34,21){$\chi_{P_4}$}
\put(30,10){\vector(0,-1){2}}
\end{picture}}
}
\end{picture}
%<<<<<<<<<<<<<<<<<<<<<<<<<<<<<<<<<<<<<<<<<<<<<<<<<<<
-
%>>>>>>>>>>>>>>>>>>>>>>>>>>>>>>>>>>>>>>>>>>>>>>>>>>>
\begin{picture}(55,30)(0,0)
\put(0,-8){
%\put(0,0){\framebox(65,40){}}
%%%%%%%%%%%%%%%%%%%%%%%%%%%%%% four Gammas
\put(20,10){\begin{picture}(40,20)(0,0)
\put(0,15){\line(1,0){5}}
\put(0,5){\vector(1,0){10}}
\put(15,15){\vector(-1,0){10}}
\put(15,5){\line(-1,0){5}}
\end{picture}}
\put(-5,10){\begin{picture}(40,20)(0,0)
\put(40,10){\oval(30,10)[r]}
\put(42,2){$\bullet$}
\put(38,-5){${\Gamma_{\hspace{-.08cm}A}}_{P_2}$}
\put(42,13){$\bullet$}
\put(38,21){${\Gamma_{\hspace{-.08cm}A}}_{P_1}$}
\put(55,10){\vector(0,1){2}}
\end{picture}}
\put(-25,10){\begin{picture}(40,20)(0,0)
\put(45,10){\oval(30,10)[l]}
\put(38,2){$\bullet$}
\put(34,-5){${\Gamma_{\hspace{-.08cm}A}}_{P_3}$}
\put(38,13){$\bullet$}
\put(34,21){$\chi_{P_4}$}
\put(30,10){\vector(0,-1){2}}
\end{picture}}
}
\end{picture}
%<<<<<<<<<<<<<<<<<<<<<<<<<<<<<<<<<<<<<<<<<<<<<<<<<<<
\ ,
\label{ciclic 2}
\end{equation}
plus, with factor $(2 \, i \, n_\pi)^{-2}$ , (4 permutations)
%
%CYCLIC 3
%
\begin{equation}
%>>>>>>>>>>>>>>>>>>>>>>>>>>>>>>>>>>>>>>>>>>>>>>
\begin{picture}(55,55)(0,0)
\put(0,-8){
%\put(0,0){\framebox(55,65){}}
%%%%%%%%%%%%%%%%%%%%%%%%%%%%%% box with two gammas
\put(16,13){\begin{picture}(20,40)(0,0)
\put(5,0){\line(0,1){5}}
\put(5,15){\vector(0,-1){10}}
\put(15,15){\line(0,-1){5}}
\put(15,0){\vector(0,1){10}}
\put(0,15){\framebox(20,10){}}
\put(5,25){\line(0,1){5}}
\put(5,40){\vector(0,-1){10}}
\put(15,40){\line(0,-1){5}}
\put(15,25){\vector(0,1){10}}
\end{picture}}
\put(16,8){\begin{picture}(20,40)(0,0)
\put(10,40){\oval(10,26)[t]}
\put(2,42){$\bullet$}
\put(-13,48){$\chi_{P_4}$}
\put(13,42){$\bullet$}
\put(18,48){$\chi_{P_1}$}
\put(10,53){\vector(-1,0){2}}
\end{picture}}
\put(16,-27){\begin{picture}(20,40)(0,0)
\put(10,45){\oval(10,26)[b]}
\put(2,38){$\bullet$}
\put(-15,34){${\Gamma_{\hspace{-.08cm}A}}_{P_3}$}
\put(13,38){$\bullet$}
\put(18,34){${\Gamma_{\hspace{-.08cm}A}}_{P_2}$}
\put(10,32){\vector(1,0){2}}
\end{picture}}
}
\end{picture}
%<<<<<<<<<<<<<<<<<<<<<<<<<<<<<<<<<<<<<<<<<<<<<<<<<<<
+
%>>>>>>>>>>>>>>>>>>>>>>>>>>>>>>>>>>>>>>>>>>>>>>
\begin{picture}(80,30)(0,0)
\put(0,-8){
%\put(0,0){\framebox(90,40){}}
%%%%%%%%%%%%%%%%%%%%%%%%%%%%%% box with two gammas
\put(20,10){\begin{picture}(40,20)(0,0)
\put(0,15){\line(1,0){5}}
\put(0,5){\vector(1,0){10}}
\put(15,15){\vector(-1,0){10}}
\put(15,5){\line(-1,0){5}}
\put(15,0){\framebox(10,20){}}
\put(25,15){\line(1,0){5}}
\put(25,5){\vector(1,0){10}}
\put(40,15){\vector(-1,0){10}}
\put(40,5){\line(-1,0){5}}
\end{picture}}
\put(20,10){\begin{picture}(40,20)(0,0)
\put(40,10){\oval(30,10)[r]}
\put(42,2){$\bullet$}
\put(38,-5){${\Gamma_{\hspace{-.08cm}A}}_{P_2}$}
\put(42,13){$\bullet$}
\put(38,21){$\chi_{P_1}$}
\put(55,10){\vector(0,1){2}}
\end{picture}}
\put(-25,10){\begin{picture}(40,20)(0,0)
\put(45,10){\oval(30,10)[l]}
\put(38,2){$\bullet$}
\put(34,-5){${\Gamma_{\hspace{-.08cm}A}}_{P_3}$}
\put(38,13){$\bullet$}
\put(34,21){$\chi_{P_4}$}
\put(30,10){\vector(0,-1){2}}
\end{picture}}
}
\end{picture}
%<<<<<<<<<<<<<<<<<<<<<<<<<<<<<<<<<<<<<<<<<<<<<<<<<<<
-
%>>>>>>>>>>>>>>>>>>>>>>>>>>>>>>>>>>>>>>>>>>>>>>>>>>>
\begin{picture}(55,30)(0,0)
\put(0,-8){
%\put(0,0){\framebox(65,40){}}
%%%%%%%%%%%%%%%%%%%%%%%%%%%%%% four Gammas
\put(20,10){\begin{picture}(40,20)(0,0)
\put(0,15){\line(1,0){5}}
\put(0,5){\vector(1,0){10}}
\put(15,15){\vector(-1,0){10}}
\put(15,5){\line(-1,0){5}}
\end{picture}}
\put(-5,10){\begin{picture}(40,20)(0,0)
\put(40,10){\oval(30,10)[r]}
\put(42,2){$\bullet$}
\put(38,-5){${\Gamma_{\hspace{-.08cm}A}}_{P_2}$}
\put(42,13){$\bullet$}
\put(38,21){$\chi_{P_1}$}
\put(55,10){\vector(0,1){2}}
\end{picture}}
\put(-25,10){\begin{picture}(40,20)(0,0)
\put(45,10){\oval(30,10)[l]}
\put(38,2){$\bullet$}
\put(34,-5){${\Gamma_{\hspace{-.08cm}A}}_{P_3}$}
\put(38,13){$\bullet$}
\put(34,21){$\chi_{P_4}$}
\put(30,10){\vector(0,-1){2}}
\end{picture}}
}
\end{picture}
%<<<<<<<<<<<<<<<<<<<<<<<<<<<<<<<<<<<<<<<<<<<<<<<<<<<
\ ,
\label{ciclic 3}
\end{equation}
plus, with factor $(2 \, i \, n_\pi)^{-2}$ , (2 permutations)
%
%CYCLIC 4
%
\begin{equation}
%>>>>>>>>>>>>>>>>>>>>>>>>>>>>>>>>>>>>>>>>>>>>>>
\begin{picture}(55,55)(0,0)
\put(0,-8){
%\put(0,0){\framebox(55,65){}}
%%%%%%%%%%%%%%%%%%%%%%%%%%%%%% box with two gammas
\put(16,13){\begin{picture}(20,40)(0,0)
\put(5,0){\line(0,1){5}}
\put(5,15){\vector(0,-1){10}}
\put(15,15){\line(0,-1){5}}
\put(15,0){\vector(0,1){10}}
\put(0,15){\framebox(20,10){}}
\put(5,25){\line(0,1){5}}
\put(5,40){\vector(0,-1){10}}
\put(15,40){\line(0,-1){5}}
\put(15,25){\vector(0,1){10}}
\end{picture}}
\put(16,8){\begin{picture}(20,40)(0,0)
\put(10,40){\oval(10,26)[t]}
\put(2,42){$\bullet$}
\put(-13,48){$\chi_{P_4}$}
\put(13,42){$\bullet$}
\put(18,48){${\Gamma_{\hspace{-.08cm}A}}_{P_1}$}
\put(10,53){\vector(-1,0){2}}
\end{picture}}
\put(16,-27){\begin{picture}(20,40)(0,0)
\put(10,45){\oval(10,26)[b]}
\put(2,38){$\bullet$}
\put(-15,34){${\Gamma_{\hspace{-.08cm}A}}_{P_3}$}
\put(13,38){$\bullet$}
\put(18,34){$\chi_{P_2}$}
\put(10,32){\vector(1,0){2}}
\end{picture}}
}
\end{picture}
%<<<<<<<<<<<<<<<<<<<<<<<<<<<<<<<<<<<<<<<<<<<<<<<<<<<
+
%>>>>>>>>>>>>>>>>>>>>>>>>>>>>>>>>>>>>>>>>>>>>>>
\begin{picture}(80,30)(0,0)
\put(0,-8){
%\put(0,0){\framebox(90,40){}}
%%%%%%%%%%%%%%%%%%%%%%%%%%%%%% box with two gammas
\put(20,10){\begin{picture}(40,20)(0,0)
\put(0,15){\line(1,0){5}}
\put(0,5){\vector(1,0){10}}
\put(15,15){\vector(-1,0){10}}
\put(15,5){\line(-1,0){5}}
\put(15,0){\framebox(10,20){}}
\put(25,15){\line(1,0){5}}
\put(25,5){\vector(1,0){10}}
\put(40,15){\vector(-1,0){10}}
\put(40,5){\line(-1,0){5}}
\end{picture}}
\put(20,10){\begin{picture}(40,20)(0,0)
\put(40,10){\oval(30,10)[r]}
\put(42,2){$\bullet$}
\put(38,-5){$\chi_{P_2}$}
\put(42,13){$\bullet$}
\put(38,21){${\Gamma_{\hspace{-.08cm}A}}_{P_1}$}
\put(55,10){\vector(0,1){2}}
\end{picture}}
\put(-25,10){\begin{picture}(40,20)(0,0)
\put(45,10){\oval(30,10)[l]}
\put(38,2){$\bullet$}
\put(34,-5){${\Gamma_{\hspace{-.08cm}A}}_{P_3}$}
\put(38,13){$\bullet$}
\put(34,21){$\chi_{P_4}$}
\put(30,10){\vector(0,-1){2}}
\end{picture}}
}
\end{picture}
%<<<<<<<<<<<<<<<<<<<<<<<<<<<<<<<<<<<<<<<<<<<<<<<<<<<
-
%>>>>>>>>>>>>>>>>>>>>>>>>>>>>>>>>>>>>>>>>>>>>>>>>>>>
\begin{picture}(55,30)(0,0)
\put(0,-8){
%\put(0,0){\framebox(65,40){}}
%%%%%%%%%%%%%%%%%%%%%%%%%%%%%% four Gammas
\put(20,10){\begin{picture}(40,20)(0,0)
\put(0,15){\line(1,0){5}}
\put(0,5){\vector(1,0){10}}
\put(15,15){\vector(-1,0){10}}
\put(15,5){\line(-1,0){5}}
\end{picture}}
\put(-5,10){\begin{picture}(40,20)(0,0)
\put(40,10){\oval(30,10)[r]}
\put(42,2){$\bullet$}
\put(38,-5){$\chi_{P_2}$}
\put(42,13){$\bullet$}
\put(38,21){${\Gamma_{\hspace{-.08cm}A}}_{P_1}$}
\put(55,10){\vector(0,1){2}}
\end{picture}}
\put(-25,10){\begin{picture}(40,20)(0,0)
\put(45,10){\oval(30,10)[l]}
\put(38,2){$\bullet$}
\put(34,-5){${\Gamma_{\hspace{-.08cm}A}}_{P_3}$}
\put(38,13){$\bullet$}
\put(34,21){$\chi_{P_4}$}
\put(30,10){\vector(0,-1){2}}
\end{picture}}
}
\end{picture}
%<<<<<<<<<<<<<<<<<<<<<<<<<<<<<<<<<<<<<<<<<<<<<<<<<<<
\ .
\label{ciclic 4}
\end{equation}

\par
To repeat the main step of eq. (\ref{step}), replacing one
of the two ${\Gamma_{\hspace{-.08cm}A}}_{P}$ by a 
$\chi_{_P}$, 
it is convenient to include (or remove) a ladder in 
the pion vertex $\chi_{_P}$. 
In the neighborhood of a boundstate 
$b$ pole $M^2_b$ in the external momentum $P^2$,
\FL
\begin{equation}
%>>>>>>>>>>>>>>>>>>>>>>>>>>>>>>>>>>>>>>>>>>>>>>
\begin{picture}(47,20)(0,0)
\put(0,-8){
%\put(0,0){\framebox(50,30){}}
%
\put(1,5){
\begin{picture}(40,20)(0,0)
%%%%%%%%%%%%%%%%%%%%%%%%%%%%%% box
\put(0,15){\line(1,0){5}}
\put(0,5){\vector(1,0){10}}
\put(15,15){\vector(-1,0){10}}
\put(15,5){\line(-1,0){5}}
\put(15,0){\framebox(10,20){}}
\put(25,15){\line(1,0){5}}
\put(25,5){\vector(1,0){10}}
\put(40,15){\vector(-1,0){10}}
\put(40,5){\line(-1,0){5}}
\end{picture}}
}
\end{picture}
%<<<<<<<<<<<<<<<<<<<<<<<<<<<<<<<<<<<<<<<<<<<
=
%>>>>>>>>>>>>>>>>>>>>>>>>>>>>>>>>>>>>>>>>>>>>>>>>>>>
\begin{picture}(42,15)(0,0)
\put(-5,-10){
%\put(0,0){\framebox(50,30){}}
%%%%%%%%%%%%%%%%%%%%%%%%%%%%%% Ket
%
\put(-20,5){
\begin{picture}(40,20)(0,0)
%%%%%%%%%%%%%%%%%%%%%%%%%%%%%% propagators
\put(25,15){\line(1,0){5}}
\put(25,5){\vector(1,0){10}}
\put(40,15){\vector(-1,0){10}}
\put(40,5){\line(-1,0){5}}
\end{picture}}
\put(-20,5){
\begin{picture}(40,20)(0,0)
%%%%%%%%%%%%%%%%%%%%%%%%%%%%%% half closed right
\put(40,10){\oval(10,10)[r]}
\put(43,8){$\bullet$}
\put(48,10){${\chi_b}_{_{P}}$}
\end{picture}}
}
\end{picture}
%<<<<<<<<<<<<<<<<<<<<<<<<<<<<<<<<<<<<<<<<<<<<<<<<<<<
\
{ i \over P^2-M_b^2 +i \epsilon} 
\
%>>>>>>>>>>>>>>>>>>>>>>>>>>>>>>>>>>>>>>>>>>>>>>>>>>>
\begin{picture}(51,15)(0,0)
\put(-4,-10){
%\put(0,0){\framebox(55,30){}}
%%%%%%%%%%%%%%%%%%%%%%%%%%%%%% Bra
%
\put(32,5){
\begin{picture}(40,20)(0,0)
%%%%%%%%%%%%%%%%%%%%%%%%%%%%%% propagators 
\put(0,15){\line(1,0){5}}
\put(0,5){\vector(1,0){10}}
\put(15,15){\vector(-1,0){10}}
\put(15,5){\line(-1,0){5}}
\end{picture}}
\put(-13,5){
\begin{picture}(40,20)(0,0)
%%%%%%%%%%%%%%%%%%%%%%%%%%%%%% half closed left
\put(45,10){\oval(10,10)[l]}
\put(37,8){$\bullet$}
\put(15,10){${\chi_b}_{_{-P}} $}
\end{picture}}
}
\end{picture}
%<<<<<<<<<<<<<<<<<<<<<<<<<<<<<<<<<<<<<<<<<<<<<<<<<<<
\label{ladder pole}
\end{equation}
multiplying the right of both sides with the vertex $\chi$, 
%
%POLE PION
%
\begin{eqnarray}
&
%>>>>>>>>>>>>>>>>>>>>>>>>>>>>>>>>>>>>>>>>>>>>>>>>>>>>>>>>>>>>>>
\begin{picture}(35,15)(0,0) 
\put(0,-10){
%\put(0,0){\framebox(35,30){}}
%
\put(0,5){                         
\begin{picture}(15,20)(0,0)
%%%%%%%%%%%%%%%%%%%%%%%%%%%%% right propagators
\put(0,15){\line(1,0){5}}
\put(0,5){\vector(1,0){10}}
\put(15,15){\vector(-1,0){10}}
\put(15,5){\line(-1,0){5}}
\end{picture}}
\put(15,5){
\begin{picture}(10,20)(0,0)
%%%%%%%%%%%%%%%%%%%%%%%%%%%% WF
\put(0,10){\oval(10,10)[r]}
\put(3,7.5){$\bullet$}
\put(8,10){$\chi$}
\end{picture}}
}
\end{picture}
%<<<<<<<<<<<<<<<<<<<<<<<<<<<<<<<<<<<<<<<<<<<<<<<<<<<<<<<<<<<<<<
=
%>>>>>>>>>>>>>>>>>>>>>>>>>>>>>>>>>>>>>>>>>>>>>>>>>>>>>>>>>>>>>>
\begin{picture}(60,20)(0,0) 
\put(0,-10){
%\put(0,0){\framebox(60,30){}}
%
\put(0,5){                         
\begin{picture}(25,20)(0,0)
%%%%%%%%%%%%%%%%%%%%%%%%%%%%% left propagators + box
\put(0,15){\line(1,0){5}}
\put(0,5){\vector(1,0){10}}
\put(15,15){\vector(-1,0){10}}
\put(15,5){\line(-1,0){5}}
\put(15,0){\framebox(10,20){}}
\end{picture}}
\put(25,5){                         
\begin{picture}(15,20)(0,0)
%%%%%%%%%%%%%%%%%%%%%%%%%%%%% right propagators
\put(0,15){\line(1,0){5}}
\put(0,5){\vector(1,0){10}}
\put(15,15){\vector(-1,0){10}}
\put(15,5){\line(-1,0){5}}
\end{picture}}
\put(40,5){
\begin{picture}(10,20)(0,0)
%%%%%%%%%%%%%%%%%%%%%%%%%%%% WF
\put(0,10){\oval(10,10)[r]}
\put(3,7.5){$\bullet$}
\put(8,10){$\chi$}
\end{picture}}
}
\end{picture}
%<<<<<<<<<<<<<<<<<<<<<<<<<<<<<<<<<<<<<<<<<<<<<<<<<<<<<<<<<<<<<
\ { P^2-M_\pi^2 \over i \, {\cal I} } 
&
\ ,
\nonumber \\
&
{\cal I} =
%>>>>>>>>>>>>>>>>>>>>>>>>>>>>>>>>>>>>>>>>>>>>>>>>>>>>>>>>>>>>>>
\begin{picture}(50,20)(0,0) 
\put(0,-10){
%\put(0,0){\framebox(50,30){}}
%
\put(0,5){
\begin{picture}(15,20)(0,0)
%%%%%%%%%%%%%%%%%%%%%%%%%%%% WF
\put(15,10){\oval(10,10)[l]}
\put(7,7.5){$\bullet$}
\put(0,10){$\chi$}
\end{picture}}
\put(15,5){                         
\begin{picture}(15,20)(0,0)
%%%%%%%%%%%%%%%%%%%%%%%%%%%%% right propagators
\put(0,15){\line(1,0){5}}
\put(0,5){\vector(1,0){10}}
\put(15,15){\vector(-1,0){10}}
\put(15,5){\line(-1,0){5}}
\end{picture}}
\put(30,5){
\begin{picture}(10,20)(0,0)
%%%%%%%%%%%%%%%%%%%%%%%%%%%% WF
\put(0,10){\oval(10,10)[r]}
\put(3,7.5){$\bullet$}
\put(8,10){$\chi$}
\end{picture}}
}
\end{picture}
%<<<<<<<<<<<<<<<<<<<<<<<<<<<<<<<<<<<<<<<<<<<<<<<<<<<<<<<<<<<<<<
&
\ ,
\label{pole pion}
\end{eqnarray}
where ${\cal I}$ is finite and proportional 
to the square of the scale of the interaction,
nevertheless it will factorize from the results.
Moreover the off mass shell Bethe Salpeter 
equation is derived replacing 
eq. (\ref{pole pion}) in eq. (\ref{ladder}), 
\begin{equation}
%>>>>>>>>>>>>>>>>>>>>>>>>>>>>>>>>>>>>>>>>>>>>>>>>>>>>>>>>>>>>>>
\begin{picture}(20,20)(0,0) 
\put(0,0){
%\put(0,0){\framebox(60,30){}}
%
\put(0,-5){
\begin{picture}(10,20)(0,0)
%%%%%%%%%%%%%%%%%%%%%%%%%%%% WF
\put(3,7.5){$\bullet$}
\put(8,10){$\chi_{_P}$}
\put(0,10){\oval(10,10)[r]}
\end{picture}}
}
\end{picture}
%<<<<<<<<<<<<<<<<<<<<<<<<<<<<<<<<<<<<<<<<<<<<<<<<<<<<<<<<<<<<<<
=
%>>>>>>>>>>>>>>>>>>>>>>>>>>>>>>>>>>>>>>>>>>>>>>>>>>>>>>>>>>>>>>
\begin{picture}(40,15)(0,0) 
\put(0,-10){
%\put(0,0){\framebox(35,30){}}
%
\put(0,5){                         
\begin{picture}(15,20)(0,0)
%%%%%%%%%%%%%%%%%%%%%%%%%%%%% potential + right propagators
\multiput(0,3)(0,2){6}{$\cdot$}
\put(0,15){\line(1,0){5}}
\put(0,5){\vector(1,0){10}}
\put(15,15){\vector(-1,0){10}}
\put(15,5){\line(-1,0){5}}
\end{picture}}
\put(15,5){
\begin{picture}(10,20)(0,0)
%%%%%%%%%%%%%%%%%%%%%%%%%%%% WF
\put(0,10){\oval(10,10)[r]}
\put(3,7.5){$\bullet$}
\put(8,10){$\chi_{_P}$}
\end{picture}}
}
\end{picture}
%<<<<<<<<<<<<<<<<<<<<<<<<<<<<<<<<<<<<<<<<<<<<<<<<<<<<<<<<<<<<<<
\left( 1 - { P^2-M_\pi^2 \over i \, {\cal I} } \right)^{-1}
\ .
\label{off mass shell}
\end{equation}
The off mass shell correction only contributes to the Bethe
Salpeter equation (\ref{off mass shell}) at the order $P^2$ 
and $M^2$. Therefore, up to the first order in the external
$P^\mu$, the vertex $\chi_{_P}$ is 
{\em formally the same function } of $P^\mu$,
both for mass shell and for off mass shell pions. 
\par
The diagrams eventually simplify into traces 
related to the pion decay relation
\cite{Pagels},
\begin{equation}
tr\{ (S \chi S)_{_{P}}
\gamma^\mu  \gamma^5 \} = 2 f_\pi P^\mu \ ,
\label{decay}
\end{equation}
which measures $f_\pi$ in the electroweak decay of pions 
on the mass shell. Because off mass shell vertices $\chi_{_P}$
are formally the same as the mass shell ones, 
eq. (\ref{decay}) is also correct for $P^2 \neq M^2$. This is experimentally confirmed
with the Goldberger Treiman relation 
\cite{Goldberger}.  
Another useful trace is directly extracted, substituting the 
spectral decomposition of the ladder (\ref{ladder pole}) 
in eq. (\ref{second}),
\begin{equation}
tr\{ (S \chi S)_{_{-P}}
{\gamma_{\hspace{-.08cm}A}}_{_{P}} \}
= 2 \, n_\pi ( P^2 -M_\pi^2 ) \ .
\label{p2-m2}
\end{equation}
A detailed momentum analysis of eqs (\ref{decay},\ref{p2-m2})
shows that the norm $ n_\pi = i\, f_\pi$ and 
produces the desired traces,
\FL
\begin{eqnarray}
tr\{ (S \chi S)_{_{P_1}}
{\gamma_{\hspace{-.08cm}A}}_{P_2} \}
&=&
 -2 \, n_\pi ( P_1 \cdot P_2 +M_\pi^2 ) \ ,
\label{condensed}
\\
-m tr\{ S \}
&=&
 n_\pi^2 \, M_\pi^2 \ ,
\label{GMOR}
\end{eqnarray}
including the Gell-Mann Oakes and Renner relation
\cite{Gell-Mann}.

\par
This completes the necessary steps to compute the loop (\ref{loop}).
The diagrams of eq. (\ref{ciclic 1}) are trivial to compute,
completing eq. (\ref{high Ward}) with the 
Gell-Mann Oakes and Renner relation (\ref{GMOR}).
The three other cases (\ref{ciclic 2},\ref{ciclic 3},\ref{ciclic 4})
are now  straightforwardly computed with the techniques that lead to  
eq. (\ref{box with ladder}), replacing eq. (\ref{second}) by
eq. (\ref{pole pion}). 
It is clear that when we replace a $\gamma_A$
by $ \chi \, (P^2_j -M_\pi^2) /i \, I$,  
the order in $P_j$ increases. Therefore the trace 
that replaces the term with a pair of 
$\{\gamma_A, \gamma_5\}$ in eq. (\ref{box with ladder}) 
now vanishes up to the order of $P_i P_j$ and of $M_\pi^2$.
The trace that replaces the term of order $\{\gamma_A, \gamma_5\}$ 
in eq. (\ref{box with ladder}) is now computed with the trace (\ref{condensed}).
Summing the total contributions to the Feynman loop of eq. (\ref{loop}) we get,
\begin{eqnarray}
&
+3 \left({ 1 \over 2 i\, n_\pi}\right)^4 
( -8 \, i \, n_\pi^2 M_\pi^2)
&
\nonumber \\
&
-2 \left({ 1 \over 2 i\, n_\pi}\right)^3 \sum_{ 4 \, perm.} 
\, 2 \, n_\pi ( P_1^2 + P_1 \cdot P_3 -2 \, M_\pi^2 )
&
\nonumber \\
&
+ \left({ 1 \over 2 i\, n_\pi}\right)^2 \sum_{ 4 \, perm.}
{P_1^2 + P_2^2 +P_1 \cdot P_2 
+P_1 \cdot P_3 +P_2 \cdot P_4 -2 \, M_\pi^2 \over i }
&
\nonumber \\
&
+ \left({1 \over 2 i\, n_\pi}\right)^2 \sum_{ 2 \, perm.}
{P_1^2 + P_3^2 -2 \, M_\pi^2 \over i }
&
\nonumber \\
&
={i \over 4 f_\pi^2} \left[
(P_1+P_2)^2+(P_1+P_4)^2
 -2 \, M_\pi^2 \right]
\ ,
&
\label{loop result}
\end{eqnarray}
where the conservation
$P_1+P_2+P_3+P_4=0$ of momentum 
was used to simplify the result.

\par
We finally compute the $\pi-\pi$ scattering matrix T.
We simply have to match the external pions, $i1$ and $i2$ 
incoming and $o1$ and $o2$
outgoing, with the four pion vertex that
we just computed in eq.(\ref{loop result}),
\begin{equation}
%>>>>>>>>>>>>>>>>>>>>>>>>>>>>>>>>>>>>>>>>>>>>>>>>>>>
\begin{picture}(200,30)(0,0)
\put(0,-8){
%\put(0,0){\framebox(200,40){}}
%%%%%%%%%%%%%%%%%%%%%%%%%%%%%% four momenta
%
\put(90,10){
\begin{picture}(40,20)(0,0)
%%%%%%%%%%%%%%%%%%%%%%%%%%%%%% lines
\put(10,10){\line(1,1){15}}
\put(10,10){\line(1,-1){15}}
\put(10,10){\line(-1,1){15}}
\put(10,10){\line(-1,-1){15}}
\put(25,25){\vector(-1,-1){10}}
\put(-5,-5){\vector(1,1){10}}
\put(-5,25){\vector(1,-1){10}}
\put(25,-5){\vector(-1,1){10}}
\end{picture}}
\put(70,10){
\begin{picture}(40,20)(0,0)
%%%%%%%%%%%%%%%%%%%%%%%%%%%%% circle and momenta
\put(30,10){\circle*{10}}
\put(2,20){$P_4$}
\put(48,20){$P_1$}
\put(48,-6){$P_2$}
\put(2,-6){$P_3$}
\end{picture}}
\put(0,10){
\begin{picture}(40,20)(0,0)
%%%%%%%%%%%%%%%%%%%%%%%%%%%%% out left
\put(2,20){$\leftarrow \ q_{o1},\tau_{o1}$}
\put(10,7){OUT}
\put(2,-6){$\leftarrow \ q_{o2},\tau_{o2}$}
\end{picture}}
\put(150,10){
\begin{picture}(40,20)(0,0)
%%%%%%%%%%%%%%%%%%%%%%%%%%%%% in right
\put(2,20){$q_{i1},\tau_{i1} \ \leftarrow $}
\put(25,7){IN}
\put(2,-6){$q_{i2},\tau_{i2} \ \leftarrow $}
\end{picture}}
}
\end{picture}
%<<<<<<<<<<<<<<<<<<<<<<<<<<<<<<<<<<<<<<<<<<<<<<<<<<<
\label{6 combinations}
\end{equation}
where the loop (\ref{loop}) is represented by the full circle. 
The loop is 
topologically invariant for cyclic permutations of 
$P_1,P_2,P_3$ and $P_4$. 
To remove double counting we fix
one match, say  $P_1=q_{i1}$. Then there are
six different combinations of the remaining
external legs. 

\par
In what concerns color, all the combinations are
identical because the pion is a color singlet, and the color
factor is appropriately included in the definition of 
$f_\pi$, see eq. (\ref{decay}). 
In what concerns momentum, we express our 
result in the usual Mandelstam
relativistic invariant variables $s, t$ and $u$.
For instance the direct combination in eq. (\ref{6 combinations})
produces the result $ i( s + u -2 M_\pi^2)/( 4 f_\pi^2 )$.
We now introduce flavor. For simplicity it was not regarded
in the previous definitions of the vertices ${\Gamma_A}_P$
and $\chi_{_P}$. Because the pion is an isovector,
we have three different cases $I=0$, $I=1$ and $I=2$. 
The flavor contributions to the pion vertex
simply factorize from the momentum contribution,
and the different combinations only produce two classes
of flavor traces, see Table \ref{flavor}.
Summing the six possible combinations of color, spin, momentum 
and flavor traces, 
and dividing by $-i$, we finally get the $\pi-\pi$
scattering $T^I$ matrix,
\begin{eqnarray}
T^0=& 
- { 2s- M_\pi^2\over 2 f_\pi^2}& - {s+t+u-4 M_\pi^2 \over 2 f_\pi^2} \ ,
\nonumber \\
T^1=& 
- { t-u \over 2 f_\pi^2}& \ ,
\nonumber \\
T^2=&
- { -s+2M_\pi^2\over 2 f_\pi^2}& \, - {s+t+u-4 M_\pi^2 \over 2 f_\pi^2 } \ ,
\label{T matrix}
\end{eqnarray}
where $s+t+u-4 M_\pi^2$ expresses the off mass shell contribution.
We remark that eq. (\ref{T matrix}) complies with the Gasser and Leutwyler results.
\cite{Leutwyler}.
The $T^I$ matrices of eq. (\ref{T matrix}) are computed at the tree level 
(including scalar and vector s, t, and u exchange), which 
is exact up to the order of $P_i^2 P_j^2$ and of $M_\pi^2$.

\par
The $\pi-\pi$ mass shell scattering lengths $a_0^I$ 
are simply obtained from the scattering amplitudes 
$T^0,T^2$ with the Born factor of $-1 \over 16 \pi M_\pi$,
and for vanishing 3-momenta. 
The $I=1$ case is antisymmetric so the first
scattering parameter is $a_1^1$
and the corresponding factor is 
${-1 \over 16 \pi M_\pi} { 4 \over 3 (t-u)}$, 
\begin{eqnarray}
a^0_0&=& {7 \over 32 \pi}{ M_\pi \over f_\pi^2} \ ,
\nonumber \\
a^1_1&=& {1 \over 24 \pi}{ 1 \over M_\pi \, f_\pi^2} \ ,
\nonumber \\
a^2_0&=&{-1 \over 16 \pi}{ M_\pi \over f_\pi^2} \ ,
\end{eqnarray}
this is the result of the famous Weinberg theorem for 
$\pi-\pi$ scattering
\cite{Weinberg}.
Off mass shell effects are very important for the
experiments. For instance in the scattering at $\pi-\pi$ 
threshold of a $\pi$ beam with virtual $\pi^*$
provided by a nucleon target, we estimate that the off mass 
shell effects of eq. (\ref{T matrix}) decrease $a^0_0$ by a 
factor of 0.5 and increase $a^2_0$ by a factor of 1.7.

\par
We conclude that the Weinberg theorem 
for pion-pion low energy scattering is also exact 
when the pions are off the mass shell and have a finite size. 
It is remarkable that the current quark mass $m$ 
and the traces 
$ tr\{S\}, \, tr\{(S\Gamma_A)^4\}, \, tr\{(S\chi)^2\} $,
which depend on the finite scale of the interaction, say the 
string tension $\sigma$ or $\Lambda_{QCD}$,
disappear from the final result. 
Any quark model with a chirally symmetric
interaction complies with the Weinberg theorem.
This detailed proof confirms that Goldstone
bosons non only are massless but also are 
noninteracting at low energy. 
\par
I acknowledge Emilio Ribeiro for reporting 
on non-trivial Ward identities.
I also acknowledge that discussions with the collaborators of the
recent works \cite{Gastao,Goncalo,pi-pi} motivated the search 
for an analytical proof of the Weinberg theorem and 
identified the class of diagrams that should be included 
in the box of eq. (\ref{loop}). 
%
%rrrrrrrrrrrrrrrrrrrrrrrrrrrrrrrrrrrrrrrrrrrrrrrrrrrrrrrrrrrrrrrrrrrrrrrrrrrr
%rrrrrrrrrrrrrrrrrrrrrrrrrrrrrrrrrrrrrrrrrrrrrrrrrrrrrrrrrrrrrrrrrrrrrrrrrr
%rrrrrrrrrrrrrrrrrrrrrrrrrrrrrrrrrrrrrrrrrrrrrrrrrrrrrrrrrrrrrrrrrrrrrrrrrr

%
%
%ttttttttttttttttttttttttttttttttttttttttttttttttttttttttttttttttt
\begin{table}[t]
\caption{
Table of the flavor traces. 
$tr\{ \tau_{i1} \tau_{i2}\tau_{o2}^\dagger\tau_{o1}^\dagger\}$
and
$tr\{  \tau_{i1}\tau_{o2}^\dagger\tau_{i2}\tau_{o1}^\dagger\}$
are examples of the direct and exchange cases.
}
\label{flavor}
\begin{tabular}{cccc}
$I \, m_I$ & $\tau_{i1} \tau_{i2}$ & direct & exchange
\\
\tableline
$0\,0$ & $ { \vec \sigma \cdot \vec \sigma \over 2 \sqrt{3} } $ 
& ${3 \over 2}$ & $-{1 \over 2}$ 
\\
$1\,1$ & $ { \sigma_1 \sigma_2 - \sigma_2 \sigma_1 \over 2\sqrt{2} } $ 
& $1$ & $0$
\\
$2\,2$& $ \sigma^+ \, \sigma^+ $ 
& $0$ & $1$ 
\end{tabular}
\end{table}
%ttttttttttttttttttttttttttttttttttttttttttttttttttttttt
%
%
%
\end{document}